\documentclass[letterpaper]{article} 
\usepackage{aaai24}  
\usepackage{times}  
\usepackage{helvet}  
\usepackage{courier}  
\usepackage[hyphens]{url}  
\usepackage{graphicx} 
\usepackage{natbib}  
\usepackage{caption} 
\usepackage{algorithm}
\usepackage{algorithmic}

\usepackage{newfloat}
\usepackage{listings}
\DeclareCaptionStyle{ruled}{labelfont=normalfont,labelsep=colon,strut=off} 
\floatstyle{ruled}
\newfloat{listing}{tb}{lst}{}
\floatname{listing}{Listing}
\title{Label-Free Subjective Player Experience Modelling via Let's Play Videos}
\author{
    Dave Goel, Athar Mahmoudi-Nejad, Matthew Guzdial
}
\affiliations{
    Computing Science Department, Alberta Machine Intelligence Institute\\
    University of Alberta
}

\usepackage{csvsimple}

\usepackage{multicol}
\usepackage[]{multirow}
\usepackage{subcaption}
\usepackage{appendix}

\begin{document}

\maketitle

\begin{abstract}
Player Experience Modelling (PEM) is the study of AI techniques applied to modelling a player's experience within a video game. 
PEM development can be labour-intensive, requiring expert hand-authoring or specialized data collection. 
In this work, we propose a novel PEM development approach, approximating player experience from gameplay video. 
We evaluate this approach predicting affect in the game Angry Birds via a human subject study. 
We validate that our PEM can strongly correlate with  self-reported and sensor measures of affect, demonstrating the potential of this approach.
\end{abstract}

\section{Introduction}

Player Experience Modelling (PEM) is the study and use of AI techniques for the construction of computational models of player experience \cite{player-modelling-def}. These models can analyze various information, including actions taken, level of knowledge, and emotional states, to gain an understanding of the player's experience. \cite{pemapp}. 
A model that can predict a player's experience can be useful for game designers to better gauge how the player feels about their game. Additionally, it opens up the possibility of implementing AI Directors, systems that dynamically alter the difficulty of the game in real-time based on models of a player's experience. 
However, PEM construction tends to take significant developer effort, which limits their applications.


Several studies have explored the development of models to predict player experience or \emph{affect} \cite{pastpems}. We define affect to be the tension experienced by the player.
The design of player experience models (PEMs) faces a significant challenge due to the absence of an objective ground truth. 
Gameplay-based PEMs rely on in-game data to predict the player's experience. They require specific game context and are therefore only applicable to a particular game or genre. This approach relies on strong assumptions about the relationship between player actions and their resulting experience.  Subjective PEMs rely on asking the players directly about their experiences, either through free responses or questionnaires. Due to issues with self-report methodologies, they do not tend to be very reliable on their own \cite{objective}. Objective PEMs measure changes in a player's physiology through sensors. While using multiple features such as player heartbeat and electrodermal activity (EDA) can result in an accurate model, in an applied case it can be infeasible due to its reliance on sensors \cite{objective}.

This paper presents a novel and more feasible approach to developing subjective PEMs utilizing readily available Let's Play videos. Let's Play content creators often exhibit vocal modulations reflecting their affect (i.e. are they excited or not). We hypothesize that a correlation exists between the amplitude of a Let's Player's voice and their level of affect during gameplay. We collected a dataset of Let's Play videos of Angry Birds from YouTube. We trained a Convolutional Neural Network (CNN) mapping game frames to affect using the post-processed amplitude of the audio chunks as the label. To validate the proposed model, we conducted a human subject study. Participants played an open-source clone of Angry Birds while physiological signals (approximating an objective measure of affect) and afterwards survey data (serving as a subjective PEM) were collected. Finally, we compared all results (physiological, survey, and CNN output) to evaluate the model's effectiveness.

In this paper, we present the following contributions:
 \begin{itemize}
     \item A novel approach to developing an affect prediction model using Let's Play video
     \item  A comprehensive pipeline for model development
     \item The results of a human-subject study to validate our proposed model, which provide evidence of its effectiveness, when compared to self-reported subjective measures
 \end{itemize}

\section{Related Work}

\begin{figure*}[tbh]
    \centering
    \includegraphics[scale=0.65]{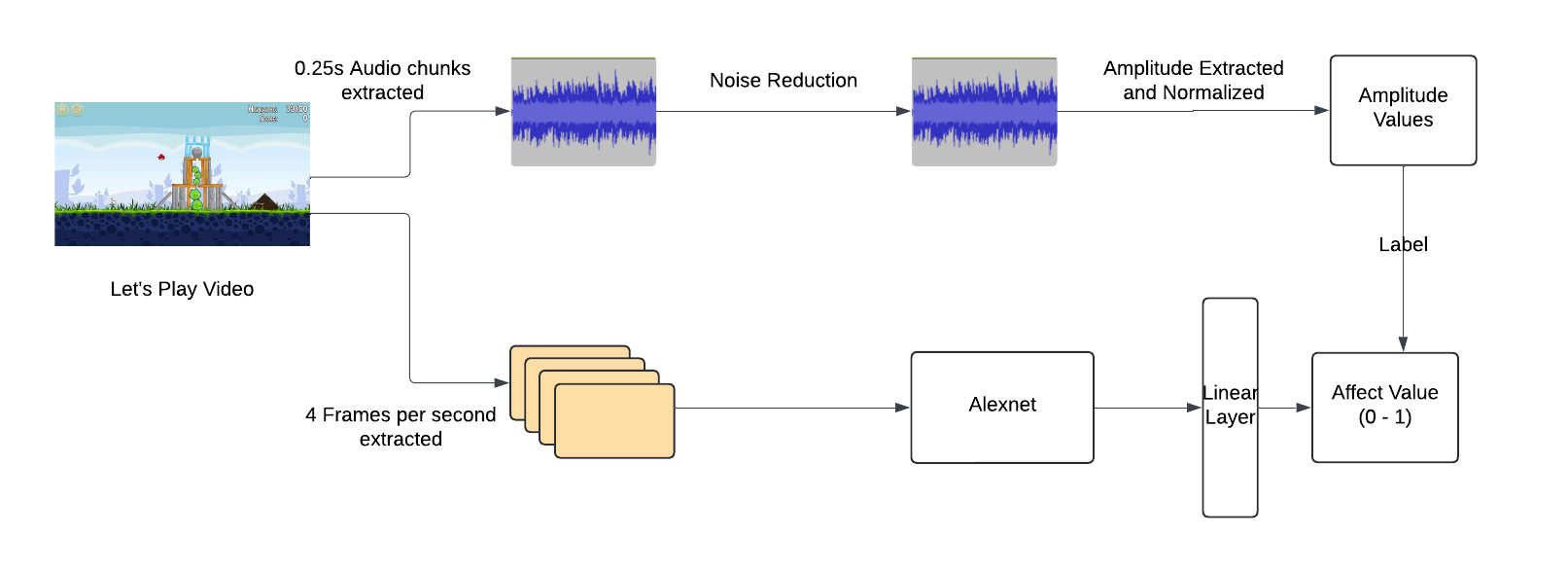}
    \caption{System Overview of our Player Experience Model pipeline}
    \label{fig:systemOverview}
\end{figure*}
Player modelling aims to understand player behavior and experience within a game. Two main approaches have been explored for player modelling. \textbf{Phsiological Signal-based Player Modelling:}  This approach attempts to measure player emotions and engagement through physiological data \cite{phymodels}. Studies have shown promise by using Electroencephalography (EEG), Electrodermal Activity (EDA), and Photoplethysmography (PPG)  to predict player affect and arousal \cite{nevermind, phy1, phy2, phy3}. Additionally, facial expression analysis \cite{facial} and eye-tracking \cite{eyetracking1, eyetracking2} have been explored. While these methods offer valuable insights, they have practical limitations as implementing sensor technology for large-scale deployments can be expensive and intrusive for players. \textbf{Gameplay Data-based Player Modelling:}  This method involves analyzing in-game interaction patterns and the context of the game to understand the player's emotions. Research has demonstrated that using this information can effectively model the player \cite{gameplay1, gameplay2}. However, a key challenge lies in interpreting the data accurately. For instance, a lack of user input could signify either deep focus on solving a puzzle or disengagement due to external distractions.

Recent studies have investigated the use of gameplay videos along with deep neural networks (DNNs) for player modelling \cite{video,makantasis2021pixels,general}. However, these approaches often depend on manually annotated data using tools like RankTrace \cite{ranktrace}, which significantly limits the size and diversity of training datasets. Zijin et al. demonstrated the ability to transfer DNN PEMs from one game to another with gameplay video but still required ground-truth labels \cite{event-extraction,gameplay-extraction}. Our proposed model provides a distinct advantage by not relying on manual annotation. This opens up the possibility of using larger and/or more diverse datasets for training, or speeding up PEM development.

\section{System Overview}


In this section, we cover our proposed approach to extract a Player Experience Model (PEM) from Let's Play video without labels.
We give an overview of the entire pipeline in Figure \ref{fig:systemOverview}. We begin by splitting a given Let's Play video into frames and audio files, such that each frame corresponds to its associated audio information. We then calculate the normalised amplitude of the audio per frame. This value is then passed through a conversion function we devised to better approximate an affect value. A CNN model is trained using this data. The frames are fed into the architecture, and the normalised amplitude values are used as labels. This model can then predict affect values by feeding in frames.

\subsection{Let's Play Videos}
For this paper, we collected a total of eight Let's Play videos from YouTube. We ensured that all the videos came from different creators, to minimize issues around the model overfitting to one creator or type of creator. We chose Let's Play videos specifically as we felt that Let's Play creators would be incentivized to overreact to game content for entertainment value, making it easier to extract affect information from the videos.

These Let's Play videos were of the game Angry Birds. We chose this game because there was an open-source version of it, Science-Birds. This allowed us to specify levels for a human subject study and to capture game logs. 

\subsection{Postprocessing}
For each video, we split the audio using Logic Pro X. Inside Logic, we applied Noise Reduction to the audio files to remove the background game music, since we are only interested in the Let's Player's voice. Using FFmpeg, we split the audio files into small chunks of length 0.25s, as this is the average human reaction time \cite{reaction}. To correspond to this, we extracted frames at a rate of four frames per second. We calculated the average normalised amplitude per audio chunk using a Python library called wavfile. 

We believe that the player's affect in the Let's Play video to be high when amplitude of the player's audio is either very high or very low. This corresponds to the player either raising their voice to react to some game event or remaining silent to concentrate. While we anticipate that most game players are regularly silent while playing, Let's Players are likely to continually speak to provide commentary and entertainment. Thus we applied a conversion function over the amplitudes. For each sample $x$, we first normalised the value to restrict its range between 0 and 1. After normalisation, we processed each sample via $cos(\pi*x)^2$, which had the effect of mapping the low and high values (0 and 1) to 1.0.  Finally, given that we are interested in general trends, we smoothed the final samples using the moving average algorithm \cite{movingavg}. 

We separately preprocessed the video frames prior to using them as training data. Individual video frames were resized to a standard dimension of 256x256 to match the expected input for AlexNet, given that we used it as the basis of our CNN model. Since a single frame might lack sufficient information, we created sequences of four consecutive frames for each data point. We grayscaled the images to preserve the input dimensions of the model and to aid in generalizability. These sequences represent a sliding window over the video, capturing temporal information.

This preprocessing pipeline resulted in two key data structures: an array containing sequences of four consecutive video frames for each data point, and an array containing the approximated affect values corresponding to each frame sequence. These processed data structures were then fed to the CNN model. We trained the model on six videos, corresponding to 26,223 training samples.   

\subsection{Model}
We used AlexNet as the basis of our model \cite{krizhevsky2017imagenet}. Given that we focused on a relatively simple game graphically and did not have a large amount of training data, we felt this model was sufficient for an initial implementation. The input dimension of the model was changed to 4*256*256, enabling it to process four consecutive frames simultaneously as a single input.
 Since the original purpose of AlexNet was image classification, we changed the last layer to output a single value between 0 and 1 instead of the original 1000 classes. This modification reflects the task at hand, which requires a continuous prediction rather than a discrete class selection. For the same reason, we used Mean Squared Error instead of Cross Entropy for our loss function. The model was trained for 20 epochs with a learning rate of 0.0001.


\section{Human Subject Study Methodology}

We chose to use a human subject study to evaluate our approach, in order to determine its effectiveness at modelling humans.
We sent out a post containing the details of our study to various discord servers local to our institution and asked potential participants to contact us over email for scheduling. 
Our study had to be in-person due to the use of sensors to provide an objective measure approximating player affect. 
Participants met student personnel in person, and were then required to fill out a consent form. 
After this, we attached a PPG sensor and an electrodermal activity (EDA) sensor to the participant's non-dominant hand to record their physiological data. We then asked them to play three levels of Angry Birds, implemented inside Science-Birds \cite{science-birds}, an open-source version of Angry Birds. 
Finally, they were asked to complete a survey designed to capture self-reported emotional experiences during the gameplay session.
This study was approved by the ethics review board at the University of Alberta, with id number Pro00140931.

\subsection{Sensors} 

This study employed PPG and electrodermal activity (EDA) to gain objective measures to approximate player affect. These non-invasive techniques are particularly well-suited for game research due to their minimal disruption to the player experience~\cite{yannakakis2016psychophysiology}. 
EDA measures autonomic changes in the skin's electrical properties, directly reflecting sympathetic arousal. This characteristic makes it the most widely used method for investigating human psychophysiology in video games~\cite{kivikangas2011review, ravaja2006phasic, ravaja2008psychophysiology, holmgaard2015multimodal, gualeni2012psychophysiology}.
PPG provides data on heart rate (HR) and heart rate variability (HRV). HRV, which indicates the variation between consecutive heartbeats, is also a valuable measure for assessing psychophysiology in games~\cite{castellar2014assessing, holmgaard2015multimodal, gualeni2012psychophysiology}.


EDA recordings have two key components: Skin Conductance Level (SCL) which reflects the overall level of arousal, exhibiting slow changes due to emotional states or sustained stimuli. Skin Conductance Response (SCR) exhibits rapid changes associated with specific stimuli or events~\cite{boucsein2012electrodermal}. 
In this study, we utilized SCL to assess participants' affect levels within the game. While SCR is also used for arousal assessment, its focus on discrete events made SCL a more suitable choice for assessing sustained arousal in our gameplay sessions.
After collecting the EDA signal, we pre-processed it to remove noise using a low-pass filter with a 3Hz cutoff frequency and a 4th-order Butterworth filter~\cite{makowski2021neurokit2}. We then extracted both SCL (a continuous signal) and SCR features (number of peaks and mean amplitude) for each participant during each level.

\begin{figure*}[tbh]
\centering
    \includegraphics[scale = 0.5]{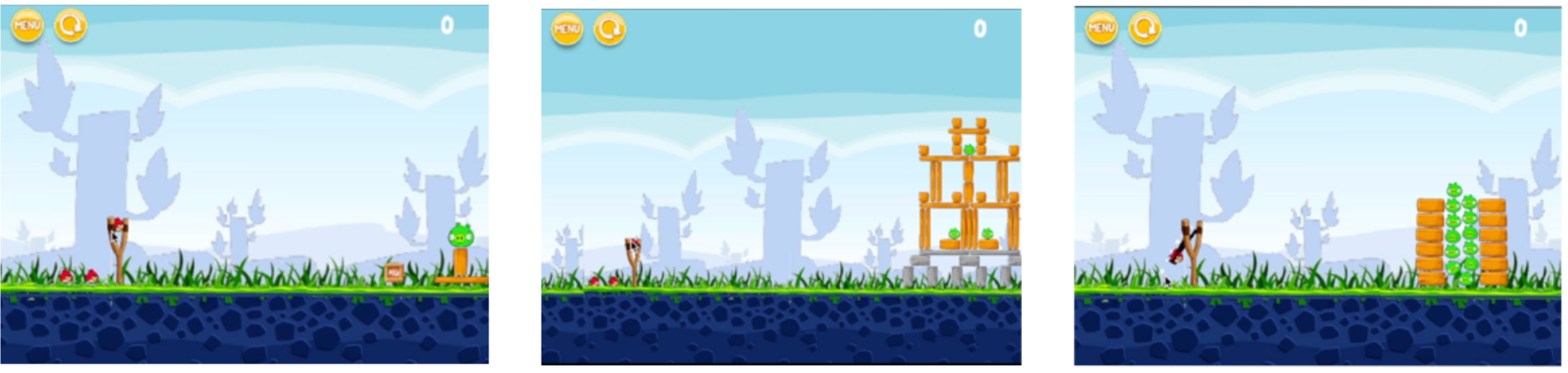}      \label{fig:sub1}
    \caption{Screenshots of levels 1, 2 and 3 from the human study}
    \end{figure*}

Similarly to EDA, HRV provides another measure for player arousal. HRV analysis can be conducted in two domains: time-domain and frequency-domain. However, frequency-domain analysis requires a minimum recording duration of 60 seconds~\cite{pham2021heart}. Since our game levels lasted less than a minute, we focused on time-domain features.
First, we de-noised the HRV signal to remove artifacts using the Butterworth filter at 0.5Hz-8Hz~\cite{elgendi2013systolic}. We then extracted time-domain features for the duration of each level for each participant, which were based on NN intervals, which are the time intervals between consecutive normal heartbeats. These features include Standard Deviation of NN intervals (SDNN), Standard Deviation of Successive NN interval Differences (SDSD), Root Mean Square of Successive NN interval Differences (RMSSD), proportion of successive NN interval differences larger than 20 milliseconds (pNN20), and proportion of successive NN interval differences larger than 50 milliseconds (pNN50), as recommended in prior work~\cite{pham2021heart}.
We used the Neurokit Python package~\cite{makowski2021neurokit2} for all physiological signal processing.

\subsection{Gameplay}

We made use of Science-birds as an open implementation of Angry Birds \cite{science-birds}.
This allowed us to track the player's telemetry and to select specific levels.
We chose three levels for participants to play, always in the same order, as shown in Figure 2.

Two of the levels were default levels already present in Science-Birds. We felt that these two levels varied in terms of difficulty to a degree that we could measure the differences in player experience. In addition, we also recreated level 1-17 from the original Angry Birds. This level was the final level of the poached eggs section, which we assume indicates the highest difficulty in this section. We chose to use the first Science-Birds level as our first level, as we felt it was a good introduction to the game. We followed this with level 1-17, as we felt this contrast would lead to a measurable change in the player's affect. Finally, we ended with the third Science-Birds level, which was also much easier than 1-17. Our hypothesis was that we could differentiate between the player's affect comparing their various measures (sensors, survey, model, etc.) between all pairs of levels. Our expectation was that players would have the highest affect in level 2, followed by level 3, and then the lowest in level 1.

We allowed participants to retry these levels until they succeeded. We did this because we felt that stopping partway through a level would not give us accurate affect measures for comparison purposes. All participants completed playing all three levels within four minutes.
We note that our model had never seen any of these levels during training or any of the data from any of the players who took part in our human subject study. This means that the model was not trained on frames from Science-Birds, which has a slightly different appearance to the true Angry Birds, requiring the model to generalize.

\subsection{Survey}

Participants filled out a survey after playing through all three levels. We include all survey questions and possible answers in the Appendix.
The survey instrument was split into five sections:

\textbf{Ground Truth for Emotional Response:} To establish a ground truth for emotional response during gameplay, participants were asked to rank the three Angry Birds levels based on their perceived affect or difficulty. These questions were based on a survey conducted for a study revolving around testing levels of the game Snakebird \cite{snakebird}. We consider this self-reported information a ground truth for their comparative emotional response.

\textbf{Emotional State Assessment:} In addition to ranking, we also evaluated the emotional state during each level using questions based on the State-Trait Anxiety Inventory (STAI) \cite{STAI}. The STAI is a validated psychological tool that measures anxiety levels, providing valuable insights into participants' emotional responses to the gameplay experience.

\textbf{Open-Ended Feedback:} An open-ended question was used to invite participants to share any additional thoughts or experiences not captured by the structured survey. 

\textbf{Gaming Experience:} To understand potential influences on emotional responses, participants were asked about their overall video game experience as well as their specific experience with Angry Birds. This information helped account for individual variations in gaming expertise and familiarity with the game mechanics. These were also adapted from the Snakebird survey \cite{snakebird}.

\textbf{Demographics:} Standard demographic questions regarding gender, age, and stimulant/depressant use within the past 24 hours were included. These factors may impact physiological responses and emotional experiences during gameplay \cite{drugs}.

\section{Results}
A total of thirteen participants took part in the human subject study. Seven of these identified as male, four as female, and two as other. Eight participants stated that they play video games daily, two stated that they play weekly, two stated that they play monthly, and one stated that they play less than once a month. When asked about how often they play Angry Birds in particular, two stated that they played frequently, seven stated that they played it occasionally, and four stated that they played it once before. Eight of the participants were in the age range 18-24, three in the age range 25-34, and two in the range 35-44.  
We believe this participant pool is well-suited for our study, as it features diverse gender representation and a high engagement with gaming, particularly among the key demographic of 18-24-year-olds. All participants also had experience with Angry Birds, indicating that they did not require additional training to play the game.

\begin{table}[h]
\centering
\begin{tabular}{|l|c|c|c|c|} \hline 

        & Level 1 & Level 2 & Level 3 & Level Ranking      \\ \hline  
Calm    & 2       & -1      & 1       & 1,3\textgreater{}2 \\ \hline  
Tense   & -2      & 1       & -1      & 2\textgreater{}1,3 \\ \hline  
Relaxed & 1       & -1      & 1       & 1\textgreater{}2   \\ \hline  
Worried & -2      & -1      & -2      & 2\textgreater{}1,3 \\ \hline  
Upset   & -2      & -2      & -2      & 2,3\textgreater{}1 \\ \hline  
Content & 1       & -1      & 1       & Inconclusive       \\ \hline 
\end{tabular}
\caption{Median values for the different experiential features across the three levels, ranging from -2 to 2. -2 = not at all, -1 = somewhat, 1 = moderately so, and 2 = very much so.}
\end{table}

\begin{table*}[]
\centering
\begin{tabular}{|c|c|c|c|c|c|}
\hline
User id                  & Output                     & Level 1     & Level 2     & Level 3      & Level Ranking                   \\ \hline
\multirow{2}{*}{User 1}  & EDA                        & 0.74 $\pm$ 0.19 & 0.22 $\pm$ 0.22 & 0.19 $\pm$ 0.05  & 1\textgreater{}2,3              \\ \cline{2-6} 
                         & Model                      & 0.79 $\pm$ 0.03 & 0.12 $\pm$ 0.02 & 0.05 $\pm$ 0.02  & 2\textgreater{}1\textgreater{}3 \\ \hline
\multirow{2}{*}{User 2}  & EDA                        & 0.91 $\pm$ 0.09 & 0.81 $\pm$ 0.12 & 0.22 $\pm$ 0.07  & 1\textgreater{}2\textgreater{}3 \\ \cline{2-6} 
                         & Model                      & 0.11 $\pm$ 0.04 & 0.12 $\pm$ 0.05 & 0.07 $\pm$ 0.03  & 1,2\textgreater{}3              \\ \hline
\multirow{2}{*}{User 3}  & EDA                        & 0.74 $\pm$ 0.12 & 0.41 $\pm$ 0.09 & 0.11 $\pm$ 0.04  & 1\textgreater{}2\textgreater{}3 \\ \cline{2-6} 
                         & Model                      & 0.14 $\pm$ 0.05 & 0.19 $\pm$ 0.05 & 0.09 $\pm$ 0.03  & 2\textgreater{}1\textgreater{}3 \\ \hline
\multirow{2}{*}{User 4}  & EDA                        & 0.64 $\pm$ 0.25 & 0.48 $\pm$ 0.22 & 0.67 $\pm$ 0.24  & 1,3\textgreater{}2              \\ \cline{2-6} 
                         & Model                      & 0.18 $\pm$ 0.03 & 0.34 $\pm$ 0.08 & 0.17 $\pm$ 0.05  & 2\textgreater{}1\textgreater{}3 \\ \hline
\multirow{2}{*}{User 5}  & EDA                        & 0.88 $\pm$ 0.1  & 0.29 $\pm$ 0.2  & 0.34 $\pm$ 0.06  & 1\textgreater{}3\textgreater{}2 \\ \cline{2-6} 
                         & Model                      & 0.23 $\pm$ 0.06 & 0.34 $\pm$ 0.05 & 0.16 $\pm$ 0.03  & 2\textgreater{}1\textgreater{}3 \\ \hline
\multirow{2}{*}{User 6}  & EDA                        & 0.63 $\pm$ 0.14 & 0.58 $\pm$ 0.23 & 0.2 $\pm$ 0.16   & 1,2\textgreater{}3              \\ \cline{2-6} 
                         & Model                      & 0.29 $\pm$ 0.04 & 0.4 $\pm$ 0.09  & 0.25 $\pm$ 0.048 & 2\textgreater{}1\textgreater{}3 \\ \hline
\multirow{2}{*}{User 7}  & EDA                        & 0.86 $\pm$ 0.09 & 0.39 $\pm$ 0.19 & 0.27 $\pm$ 0.04  & 1\textgreater{}2\textgreater{}3 \\ \cline{2-6} 
                         & Model                      & 0.26 $\pm$ 0.09 & 0.41 $\pm$ 0.1  & 0.26 $\pm$ 0.09  & 2\textgreater{}1,3              \\ \hline
\multirow{2}{*}{User 8}  & EDA                        & 0.87 $\pm$ 0.1  & 0.64 $\pm$ 0.25 & 0.39 $\pm$ 0.02  & 1\textgreater{}2\textgreater{}3 \\ \cline{2-6} 
                         & Model                      & 0.18 $\pm$ 0.05 & 0.39 $\pm$ 0.09 & 0.14 $\pm$ 0.03  & 2\textgreater{}1\textgreater{}3 \\ \hline
\multirow{2}{*}{User 9}  & EDA                        & 0.96 $\pm$ 0.02 & 0.68 $\pm$ 0.12 & 0.07 $\pm$ 0.007 & 1\textgreater{}2\textgreater{}3 \\ \cline{2-6} 
                         & Model                      & 0.29 $\pm$ 0.08 & 0.41 $\pm$ 0.09 & 0.24 $\pm$ 0.05  & 2\textgreater{}1\textgreater{}3 \\ \hline
\multirow{2}{*}{User 10} & EDA                        & 0.12 $\pm$ 0.05 & 0.39 $\pm$ 0.11 & 0.93 $\pm$ 0.06  & 3\textgreater{}2\textgreater{}1 \\ \cline{2-6} 
                         & Model                      & 0.12 $\pm$ 0.03 & 0.14 $\pm$ 0.06 & 0.07 $\pm$ 0.02  & 2,1\textgreater{}3              \\ \hline
\multirow{2}{*}{User 11} & EDA                        & 0.93 $\pm$ 0.05 & 0.51 $\pm$ 0.12 & 0.26 $\pm$ 0.04  & 1\textgreater{}2\textgreater{}3 \\ \cline{2-6} 
                         & Model                      & 0.19 $\pm$ 0.03 & 0.31 $\pm$ 0.04 & 0.18 $\pm$ 0.06  & 2\textgreater{}1,3              \\ \hline
\multirow{2}{*}{User 12} & EDA                        & 0.83 $\pm$ 0.12 & 0.35 $\pm$ 0.26 & 0.77 $\pm$ 0.15  & 1,3\textgreater{}2              \\ \cline{2-6} 
                         & Model                      & 0.21 $\pm$ 0.03 & 0.31 $\pm$ 0.08 & 0.22 $\pm$ 0.06  & 2\textgreater{}3\textgreater{}1 \\ \hline
\multirow{2}{*}{User 13} & EDA                        & 0.96 $\pm$ 0.05 & 0.45 $\pm$ 0.29 & 0.02 $\pm$ 0.01  & 1\textgreater{}2\textgreater{}3 \\ \cline{2-6} 
                         & Model                      & 0.3 $\pm$ 0.06  & 0.45 $\pm$ 0.09 & 0.29 $\pm$ 0.05  & 2\textgreater{}1\textgreater{}3 \\ \hline
\end{tabular}
\caption{Mean and standard deviation for the model and EDA output for all the users across the 3 levels.}
\end{table*}
We summarize the results of the survey in Table 1, which shows the median values for the different experiential features across all participants for each of the three levels. The last column of the table shows how the levels ranked for each of these features. These rankings were determined
by running Wilcoxon Mann-Whitney U tests ($p < 0.05$). We used this test as the Likert values were non-normal. Thus $2>1$ indicates that the experiential feature for level 2 was ranked statistically significantly higher than for level 1. As suggested by the rankings, we were able to verify that the participants felt more calm and relaxed for level 1 than level 2, and felt more tense and worried for level 2 than level 1 and 3. These rankings suggest that the players did feel that the second level stood out in terms of intensity of affect. However, it does not indicate that players could differentiate between levels 1 and 3. 

We also ran Wilcoxon Mann-Whitney U tests on the PPG data, however, they could only identify a statistically significant difference between levels 2 and 3 ($p<0.05$). Thus we do not include the PPG values in any of the following results
\begin{table}[]
\centering
\begin{tabular}{|c|c|c|c|}
\hline
        & Level 1   & Level 2  & Level 3  \\ \hline
User 1  & -0.62***  & 0.14     & 0.56***  \\ \hline
User 2  & 0.78***   & 0.04     & -0.17*   \\ \hline
User 3  & -0.87***  & -0.52*** & -0.77*** \\ \hline
User 4  & 0.42***   & -0.19**  & -0.18*   \\ \hline
User 5  & -0.708*** & 0.53***  & -0.09    \\ \hline
User 6  & -0.03     & -0.17*   & -0.35*** \\ \hline
User 7  & -0.46***  & 0.19*    & -0.25*   \\ \hline
User 8  & 0.74***   & -0.27**  & 0.62***  \\ \hline
User 9  & -0.29     & 0.03     & -0.6***  \\ \hline
User 10 & 0.63***   & 0.65***  & -0.39*** \\ \hline
User 11 & 0.21*     & -0.18*   & -0.07    \\ \hline
User 12 & -0.007    & -0.18*   & -0.46*** \\ \hline
User 13 & 0.43**    & 0.62***  & 0.23*    \\ \hline
\end{tabular}
\caption{Correlation between the model and EDA data across the 3 levels for each user. *$p<0.05$, **$p<0.005$, ***$p<0.0005$.}
\end{table}

Table 2 captures the performance of our model and EDA sensors. Columns 2-5 show the mean and standard deviation for the EDA output and model prediction across the three levels. The last column of the table shows the level rankings distribution of their EDA/Model output values, which were again calculated using the Wilcoxon Mann-Whitney U test. The EDA consistently predicts higher values for level 1, compared to level 2 and 3. This could be due to the participants feeling initially anxious with the unusual experience of having sensors placed on them and then calming down as the study progressed. In comparison, our model does consistently predict higher values for level 2 compared to the other levels. This matches the survey results and our own expectations.  
\begin{figure}
\centering
    \includegraphics[width=\linewidth]{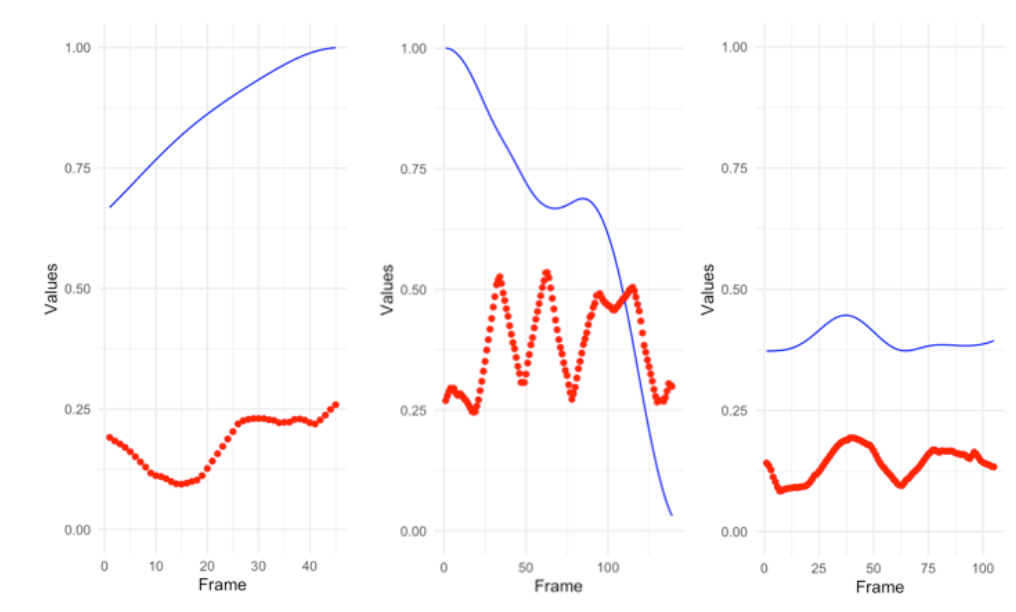}      \label{fig:sub2}
    \caption{Scatterplot depicting the model output and EDA values for User 8 for Level 1, 2 and 3.  The blue points represent the EDA output while the red points represent the model output.}
    \end{figure}
The prior results demonstrated that our model could differentiate between levels in a way that aligned with human judgement. However, a PEM model would ideally also be useful running live during a gameplay session. However, we lack any self-report data related to moment-to-moment affect. Thus, despite the potential issues with the EDA sensor values we chose to compare our model's outputs with these values. 

We ran correlation tests to compare our model's output and the EDA values for each level and participant. We used Spearman's Rho given the data does not follow a normal distribution, and give the rho values and level of significance in Table 3. We note the majority of correlations are negative, which we anticipate is due to the initial excitement at the sensors and later relaxing issue we identified above. However, this was not true for some users, like user 10. As we can see from Table 2, user 10 did not demonstrate this relaxing phenomenon. In Table 3, we can see that there was a strong, positive correlation for level 1 and level 2 for this user. We can see similar effects for other users, such as user 8, which we visualize in Figure 3. 
We take these results to indicate that our model's output could correlate positively with EDA values outside of the noise from the earlier described phenomenon. 



\section{Discussion}

We found that our model was able to predict higher affect when users self-reported higher affect. We were also able to identify instances of strong correlations between the model and EDA sensor values. However, we could not establish a satisfactory ground truth for moment-to-moment gameplay affect using the sensors.

\subsection{Limitations}
In this paper, we have proposed the label-free subjective PEM from Let's Play videos approach. As such, we made several technical and design decisions due to treating this as a proof-of-concept. The network model uses four images sampled over one second as input to the network. While we focused on this timeframe for initial exploration and real-time feasibility, we recognize the potential benefits of longer windows. 
Another issue that we ran into was that the PPG data did not end up being useful as it did not supply as rich of a signal as the EDA data and could only differentiate between levels 2 and 3. Similarly for the EDA sensor, for most of the participants, the sensor values decreased with time. As such, it is possible that these values do not capture the player's reaction to the gameplay. A possible reason for why this happened could be that Angry Birds did not cause enough ``stress'' or ``tension'' within the player, and perhaps other games that are known to be more ``stressful'' might lead to stronger sensor readings. 
Another limitation is that we did not randomize the ordering of the levels in our study. However, we chose this intentionally in order to maximize the chance for measuring changes in affect, with one low-to-high affect change (level 1 to 2) and one high-to-low affect change (level 2 to 1). Given that we were comparing our model's outputs to other measures, and not the perceived difficulty of the levels we do not think that is a major threat to validity. However, we would like to test this approach further in a follow-up study with a randomized ordering.

\subsection{Future Work}
This paper represents an initial example of our label-free subjective PEM approach, but we identify major opportunities for future work. 
Our current approach used a standard Alexnet model with a fully connected layer for regression, but we are interested in seeing how more sophisticated neural networks might perform.  Additionally, since we used a regression Convolutional Network, in the future, we would be interested in seeing how a classification paradigm performs. We also have considered applying transfer learning on a pretrained Alexnet or other Convolutional models given success with the approach in prior work \cite{event-extraction, gameplay-extraction}.
Additionally, the model could also be compared against Ranktrace annotated videos, using them as the ground truth instead of the sensors.
Beyond evaluations, we hope to create a package that can automate this pipeline, which could be used by game designers to extract a model specialised to their game. Potentially, indie studios might be able use this to get estimates of difficulty for unreleased levels for their games. We would also ideally test such a package through applications like AI directors. 


\section{Conclusions}
This paper presents a new approach to Player Experience Modelling that utilises Let's Play videos to learn an affect model. This model does not require any data annotation since it utilises the amplitude of the Let's Player's voice to approximate affect. We defined a pipeline that can be used to acquire this model, and compared it with a subjective and objective PEM by conducting a human study. Our model was successfully able to predict the overall affect for the individual levels, matching self-report measures and some objective measures.  We hope that this work will open up PEMs for broader use in academia and beyond.

\section{Acknowledgements}
This work was funded by the Canada CIFAR AI Chairs Program, Alberta Machine Intelligence Institute, and the Natural Sciences and Engineering Research Council of Canada (NSERC).

\bibliography{aaai24}

\begin{appendices}
\newpage
\clearpage
    \section{Appendix}

    Below we give the full list of survey questions, with the format or potential answers given in parentheses.

    \begin{itemize}
        \item Enter your Study ID, this will be given to you by the researcher.
    
    \textbf{Ground Truth for Emotional Response}
    
        \item Rank all the levels in terms of stress level (1 being the most stressful, and 3 being the least stressful).
    
    \textbf{Emotional State Assessment}
        \item Indicate the extent you have felt calm while exposed to the First level (Not at all/Somewhat/Moderately So/Very Much So).
        \item Indicate the extent you have felt tense while exposed to the First level (Not at all/Somewhat/Moderately So/Very Much So).
        \item Indicate the extent you have felt relaxed while exposed to the First level (Not at all/Somewhat/Moderately So/Very Much So).
        \item Indicate the extent you have felt worried while exposed to the First level (Not at all/Somewhat/Moderately So/Very Much So).
        \item Indicate the extent you have felt upset while exposed to the First level (Not at all/Somewhat/Moderately So/Very Much So).
        \item Indicate the extent you have felt content while exposed to the First level (Not at all/Somewhat/Moderately So/Very Much So).
        \item Indicate the extent you have felt calm while exposed to the Second level (Not at all/Somewhat/Moderately So/Very Much So).
        \item Indicate the extent you have felt tense while exposed to the Second level (Not at all/Somewhat/Moderately So/Very Much So).
        \item Indicate the extent you have felt relaxed while exposed to the Second level (Not at all/Somewhat/Moderately So/Very Much So).
        \item Indicate the extent you have felt worried while exposed to the Second level (Not at all/Somewhat/Moderately So/Very Much So).
        \item Indicate the extent you have felt upset while exposed to the Second level (Not at all/Somewhat/Moderately So/Very Much So).
        \item Indicate the extent you have felt content while exposed to the Second level (Not at all/Somewhat/Moderately So/Very Much So).
        \item Indicate the extent you have felt calm while exposed to the Third level (Not at all/Somewhat/Moderately So/Very Much So).
        \item Indicate the extent you have felt tense while exposed to the Third level (Not at all/Somewhat/Moderately So/Very Much So).
        \item Indicate the extent you have felt relaxed while exposed to the Third level (Not at all/Somewhat/Moderately So/Very Much So).
        \item Indicate the extent you have felt worried while exposed to the Third level (Not at all/Somewhat/Moderately So/Very Much So).
        \item Indicate the extent you have felt upset while exposed to the Third level (Not at all/Somewhat/Moderately So/Very Much So).
        \item Indicate the extent you have felt content while exposed to the Third level (Not at all/Somewhat/Moderately So/Very Much So).
    
    \textbf{Open-Ended Feedback}
        \item Were there any particular moments where you felt high stress/tension? (Optional)
        \item Were there any particular moments where you felt low stress/tension? (Optional)
        \item Any additional comments? (Optional)
    
    \textbf{Gaming Experience}
        \item How often do you play Video Games? (Less than a month/Monthly/Weekly/Daily)
        \item Do you have Prior experience with Angry Birds? (Never Played/Played it once before/Played it occasionally/Played it frequently)
    
    \textbf{Demographics}
        \item When was the last time that you used stimulants (like coffee, energy drinks, cigars) today? (more than 6 hours ago/between 2-6 hours ago/less than 2 hours ago)
        \item When was the last time you used depressants (like alcoholic drinks, marijuana) today? (more than 6 hours ago/between 2-6 hours ago/less than 2 hours ago)
        \item What age bracket do you fall under? (18-24/25-34/35-44/45-54/55-64/65 or older)
        \item Gender Identity (Male/Female/Other/Prefer not to say)
    \end{itemize}
    
    \end{appendices}

\end{document}